\documentclass{article}
\usepackage{amssymb}

\usepackage{graphicx}
\usepackage{amsmath}

\input{tcilatex}

\begin{document}

\title{FOUR DIMENSIONAL SPHERICAL SPACE-TIMES AND STRING THEORY.}
\author{H. BOUTALEB-JOUTEI$^{\text{(1)}}$ ,A.M EL GASMI$^{\text{(1,2)}}$\textit{.} \\
$^{\text{1}}$Laboratoire de physique th\'{e}orique D\'{e}partement de \\
physique facult\'{e} des Sciences de Rabat\\
Morocco$.$\\
$^{\text{2}}$D\'{e}partement de Physique \\
Facult\'{e} des Sciences Ben M'Sik Universit\'{e}\\
Hassan II-Mohammedia Casablanca\\
Morocco$.$\\
$^{1}$e-mail$:$a.elgasmi@univh2m$.$ac$.$ma}
\maketitle

\begin{abstract}
Some shortcomings in regard to our lack of conceptual understanding of
string theory are displayed and prescription to untangle them is proposed. 
\emph{String theory should be a fundamental dynamics of four dimensional
symmetric space-times}. Properties of the two dimensional equivalent action
are studied, in the hydrodynamic approximation. In the pressureless regime
it is conformal invariant. Correlations of our proposal to 't Hooft work on
quantization of black holes$^{[7]}$ and work on 2D black hole solutions
established by Witten$^{[14]}$ are pointed out as perspectives of the
present work.
\end{abstract}

\pagebreak

\bigskip

\section{Introduction.}

\bigskip During the last century, concepts, theoretical developments and
experimental achievements dealing with structure and dynamics of nature at
different energy scales have reached a sufficient level of maturity and
clarity to have a good sense of orientation on the long time survey toward
the unified theory of nature. It is claear that Whatever its formulation, it
should satisfy the following conditions:

{\small i)} unify all the known basic interactions in a quantum relativistic
frame.

{\small ii)} permit only, or almost, one structure of matter and its
interactions at any scale or epoch of our universe.

{\small iii)} to be free of non-naturalness problems, like the fine tuning
one.

{\small iv)} to be compatible with the actual standard models. In the
particle physics it must be close to the standard model of electroweak and
strong interactions at scales $\preceq $ 100 Gev$^{[1]}$. It also should
include the hot standard Big-Bang model and the inflationary paradigm$^{[2]}$%
.

On this setting, it is commonly accepted that actually string theory shows
the most likely aspects which get from a fundamental description of nature.
Mainly string theory admits Calabi-Yau solutions with gauge structures very
close to the standard model one with parameters entirely deduced from
topological considerations$^{[3]}$, its effective action is close to the
usual N=1 supergraviry near the Planck scale and the problem of
non-renormalisability of quantum gravity is naturally solved.

Theses viable aspects are certainly encouraging results in this direction
and undoubtedly the accompanied sophisticated mathematical developments
development of M-theory, duality and p-branes solutions in string theory$%
^{[4]}$, the Holographic principle$^{[5]}$ the Ads/CFT correspondence$^{[6]}$
and the developments of Hawking-'t Hooft proposal on the construction of
Back holes quantum mechanics$^{[7]}$, are certainly important advances
toward the understanding of string theory in a technical and conceptual
context and will be very useful to subsequent developments in physics, but
analysis of its building blocks and its historical occurrence show clearly
that string theory is highly intuitively founded$^{[7]}$ and a deep
understanding of its foundation is in order.

The aim of this work is to propose an alternative description of string
theory which deals with messes in our understanding of the string theory. In
the following section we consider building blocks of string theory in order
to list such messes and suggest alternative description; string theory
should be interpreted as an effective description of four dimensional
spherical space-time. In section three we analysis the global aspect of such
effective description and obtains the following scheme: \emph{String theory
with seemingly two dimensional static target space describes in fact
dynamic, including quantum effects, in four dimensional symmetric and
time-dependant geometry. }Section four deals with the \ explicit
construction of the relevant reduced action principle and study its
classical properties in particular we show that the action is conformal
invariant in the pressureless regime. In the conclusion we discuss
perspectives of the present work.

\pagebreak

\section{\protect\bigskip Alternative description of string theory.}

\bigskip

To have an explicit look over what may be loss in our understanding of the
string theory, let's first consider its building blocks. It consist of the
following:

$1)$ A target space, including space-time, evolving upon two dimensional
space, the world sheet $\sum_{g}$ with genus g and metric $h_{\alpha \beta }$%
; according to a generalized sigma model action

\qquad \qquad 
\begin{equation}
S=\sum\limits_{g=0}^{\infty }\int\limits_{\sum_{g}}d^{2}\sigma
A_{i}(Z,\partial Z)\Phi _{i}(Z)\qquad
\end{equation}
$\qquad \qquad \qquad $

A$_{i}$(Z,$\partial $Z) is a complete set of composite of target coordinates
and theirs derivations with respect to the world-sheet coordinates $\sigma
^{\alpha }$ ($\alpha =1,2)$. A$_{i}$(Z,$\partial $Z) are in one to one
correspondence with states of string theory. $\Phi _{i}(Z)$ are string
backgrounds, target sections or 2D fields coupling constants. The essential
(renormalisable) bosonic part of (1) is

\begin{eqnarray}
S &=&\frac{1}{4\pi \alpha ^{\prime }}\tint_{\sum_{g}}d^{2}\sigma \left\{ 
\sqrt{h}h^{\alpha \beta }\partial _{\alpha }X^{\mu }\partial _{\beta }X^{\nu
}G_{\mu \nu }+\varepsilon ^{\alpha \beta }\partial _{\alpha }X^{\mu
}\partial _{\beta }X^{\nu }B_{\mu \nu }\right\} +\linebreak  \notag \\
&&\alpha ^{\prime }\tint_{\sum_{g}}d^{2}\sigma \sqrt{h}\phi
(X)R.+S_{int}\qquad
\end{eqnarray}
$\alpha ^{\prime }$ is the inverse string tension, X$^{\mu }$ ($\mu
=1,.......,D)$ are space time coordinates. Backgrounds G$_{\mu \nu },$B$%
_{\mu \nu }$ and $\phi $ are respectively space-time metric, axion field and
the dilaton field. $S_{int}$ is the internal part of (2). It is usually
represented by Landau-Ginsburg model or sigma model with internal space
coordinates.

{\scriptsize 2)} The partition function Z($\phi _{i})$, which is close the
space time effective action $\Gamma (\phi _{i}),$

\begin{equation}
Z(\phi _{i})=\sum\limits_{g=0}^{\infty }e^{\rho \chi }\int\limits_{\sum_{g}} 
\left[ dh_{\alpha \beta }\right] \left[ dX^{i}\right] \left[ dZ\right] e^{%
\frac{i}{h}S}
\end{equation}
$e^{\rho }$ is the string coupling constant, $\rho $ is the constant part of
the dilaton field, $\chi $ is the $\sum_{g}$Euler characteristic; $\chi
\left( \sum_{g}\right) =2(g+1).$

{\scriptsize 3)} N-points correlation functions 
\begin{eqnarray}
\frac{\delta ^{N}\Gamma (\Phi _{1},.........,\Phi )}{\delta \Phi
_{1}......\delta \Phi _{N}} &=&\left\langle
\prod\limits_{i=1}^{N}A_{i}\right\rangle  \notag \\
&=&\sum\limits_{g=0}^{\infty }\left[ e^{(N+1)\rho \chi
}\int\limits_{\sum_{g}}\left[ dh_{\alpha \beta }\right] \left[ dX^{i}\right] %
\left[ dZ\right] \left\{ e^{\frac{i}{h}S}\prod\limits_{i=1}^{N}A_{i}\right\} %
\right]
\end{eqnarray}
with background satisfying the quantum equation of motion$\qquad \qquad
\qquad $

\begin{equation}
\frac{\delta \Gamma (\Phi _{_{i}})}{\delta \Phi _{_{i}}}=0
\end{equation}

the effective action $\Gamma (G_{_{\mu \nu }},\Phi _{_{i}}=0),$ $G_{_{\mu
\nu }}\neq \Phi _{_{i}},$ at level g=0, or the string classical equation of
motion of the metric, can be expanded in terms of sigma model perturbation
theory

\begin{equation}
R_{_{^{_{_{_{\mu \nu }}}}}}+\alpha ^{^{\prime }}\left( R_{_{\mu \nu \rho
\sigma }}\right) ^{^{2}}+.......=0\qquad
\end{equation}
where dotes represent higher perturbatives terms constructed from
derivatives and higher powers of the curvature tensor.

There are strong hints to suspect that we need an understanding of the basic
components (equations 1$\rightarrow $6) which we usually use to describe
string theory. Let is display what is actually messy with string theory:

{\small 1)} The degeneracy of string vacua$^{[8]}$. There are many
potentially viable solutions of string theory.

{\small 2)} The lack of a consistent mechanism for supersymmetrie breaking$%
^{[8]}$.

{\small 3)} The cosmological constant problem.$^{[8]}.$

{\small 4)} The recent exciting advances in string theory were limited to
strings in time independent backgrounds. They shed any light on time
dependant dynamic of the universe, essentially on its expansion and its
inflationary phase, nor on the existence of singularities in the strong
curvature regime. Moreover attempts to study sigma models with time
dependant backgrounds shows that they are too singular $^{\left[ 9\right] };$%
our understanding is limited to backgrounds which admit in asymptotic
spatial infinity a global timelike Killing vector$^{[10]}$. This condition
rules out interesting backgrounds like those which are important in cosmology%
$^{\left[ 11\right] }.$Understanding cosmological solutions in the context
of string theory is interesting both from a conceptual and from a pragmatic
point of view. We can hope that through cosmology the much desired
connection between string theory and experiment can materialize.

{\small 5)} The perturbative interpretation of Polyakov series (this
interpretation is inherent to the string concept) seems to be incompatible
with the usual connection between loops corrections and the planck constant
h. In principle at each \ topological level of the Polyakov series a two
dimensional perturbation development contains trees terms independent of h
and loops corrections which vanish at semiclassical limit h $\longrightarrow
0.$ So in the space-time sense the effective action admits two dimensional
trees contributions from higher topological terms, this suggests a new
interpretation of the Polyakov series. \ \ \ \ \ \ \ \ \ \ \ \ \ \ \ \ \ \ 

{\small 6)} In the light-cone gauge$^{\left[ 3\right] }$ the world sheet
time variable coincides with the space time variable x$^{0}$. Evolution of
spatial coordinates upon the world sheet is given by x$^{^{i}}=$x$%
^{^{i}}(\sigma ,$x$^{^{0}})$, $\sigma $ is the spatial world sheet
coordinate. When the evolution of string in space is concerned this
dependance seems natural. But if we are interested about locality in the
spatial part of the universe the $\sigma $-dependance of x$^{^{i}}$ incites
to consider $\sigma $ as a scale of the universe in equal footing than the
cosmic time x$^{^{0}},$this indicates that the natural interpretation of the
world sheet is rather a two dimensional parameter space (base space) upon
which the universe -or a part of it- and matter evolve.

{\small 7)} The last point which goes against the string concept is the fact
that actually there is no principles which support this concept or exclude
higher dimensional extension of the point particle concept.

\bigskip Recent development of M-theory, duality and p-branes solutions in
string theory$^{[4]}$, the Holographic principle$^{[5]}$ the Ads/CFT
correspondence$^{[6]}$ and the developments of Hawking-'t Hooft proposal on
the construction of Back holes quantum mechanics$^{[7]}$, are certainly
important advances toward the understanding of string theory in a technical
and conceptual context. But they are not sufficient by themselves to solve
the above listed problems.

In our point of view the \ appropriate frame to look for an outcome to these
points is the low dimensional effective description of initially higher
dimensional symmetric dynamical systems. For example when we deal with
matter evolving on homogenous isotropic four dimensional space time M$%
_{_{4}} $, gravitational equations coupled to matter reduce to a dynamical
system with one dimensional base space M$_{_{1}}.$ This reflects the
existence of differentiable fibration

\begin{eqnarray}
f_{1} &:&E\longrightarrow M_{_{4}}  \notag \\
f_{2} &:&M_{_{4}}\longrightarrow M_{_{1}}
\end{eqnarray}

The initial dynamical system corresponding to f$_{_{1}}$ is described by a G$%
_{_{1}}$ gauge invariant and general covariant action. We should get an
analog action on M$_{_{1}}$corresponding to f$_{_{2}}$of$_{_{1}}$ with $%
G=G_{_{1}}\times \frac{G_{_{ext}}}{SO(3)},$ G$_{_{1}}$and G$_{_{ext}}$ are
respectively the M$_{_{_{4}}}$ internal symmetry and the geometric one; this
action describes evolution upon M$_{_{1}}.$ So on this setting the basic
data are (E,G), gauge invariance and general covariance principle. The
fundamental law governs G-orbits evolution upon the orbit space $\frac{E}{G}%
=M_{_{d}}$, $d=dim\frac{E}{G}=dimE-dimG$, according to a G-gauge invariant
and generally covariant action

\begin{equation}
S_{_{d}}=\int\limits_{M_{d}}\left\{ \sqrt{g}\phi
R+L_{_{mat}}+L_{_{gauge}}+L_{_{top}}\right\}
\end{equation}
$\phi $ is a scalar field (the dilaton), $\int\limits_{M_{d}}\left\{
L_{_{mat}}+L_{_{gauge}}\right\} $ is the usual matter and gauge field action
and $\int\limits_{M_{d}}\left\{ L_{_{top}}\right\} $ \ is the topological
term which picks the vacuum.

The partition function corresponding to (8) is given by the polyakov like
series

\begin{equation}
Z(\phi _{_{i}})=\sum\limits_{M-topo\log ies}^{\infty }\left\{
\int\limits_{M_{d}}\left[ dg_{_{\alpha \beta }}\right] \left[ d\Psi \right] %
\left[ dZ\right] e^{\frac{i}{h}S_{_{d}}}\right\}
\end{equation}

In this way:

\bigskip {\scriptsize i)} With d=p+1 we obtain the p-brane systems$^{\left[
12\right] }.$That means ''string'' p=1 or p-branes appear as fundamental
description of nature needs physical considerations which pick the initial
data the total space and the structure group(E,G).

{\scriptsize ii)} In the d=2 case we get a string like system:

\ $\bigstar $ An action given by S$_{_{2}}$; in the trivial connection case, 
$\int\limits_{M_{d}}\left\{ L_{_{gauge}}\right\} =0$, coincides with the
string action, the target is the G-orbit space. L$_{_{top}}$ correspond to
the Kalb-Ramond topological invariant term which is usually represented by
the Wess-Zumino-Witten term when the G-orbit space is a Lie group$^{\left[ 13%
\right] }.$

\ $\bigstar $ \emph{M}$_{2}$\emph{\ is part of space time and what may be
space-time static solutions in string Sens are 4D space time and time
dependant ones}.

\ $\bigstar $ \emph{Z is the qauntum gravity partition function; if the
quantum effects do not lead to topology change, contribution to Z comes
entirely from one term in (9) corresponding to integration over
topologically equivalent metrics.}

Interests of this scheme are:

\emph{i) To describe dynamic in four dimensional time-dependant geometry we
need a string theory with two dimensional static target space, this get over
the above listed problems.}

\emph{ii)Developments of string theory show that this dynamic including
quantum effects may entirely be deduced from evolution upon the space-time
invariant part M}$_{_{2}}$\emph{\ with respect to G}$_{_{ext}}$\emph{, in
particular when M}$_{_{2}}$\emph{=S}$^{_{2}}$\emph{.}

To develop this scheme, we first need construction of the two dimensional
equivalent action of Einstein equation in four dimensional G$_{ext}$
isometric space-time and actions describing evolution of G$_{ext}$-orbits
and matter upon M$_{2}.$ Our purpose is to deal with the two first
constructions. In the following section we consider  reduction in the case G$%
_{ext}=$ $\frac{SO(3)}{U(1)}$ and the simplest one of RFW space-times .\ In
the fourth section we consider  gravitational action in the case G$_{ext}=$ $%
\frac{SO(3)}{U(1)}$and its symmetries in the thermodynamic approximation,
they include 2D conformal symmetries in the pressureless regime.\ We
conclude by discussing perspectives of the present work.

\section{Fundamental evolution on four dimensional space-times.}

Four dimensional isotropic and homogeneous space-times admit a six
dimensional isometric group G with isotropic subgroup SO(3). This implies
that \ the four dimensional space-time M$_{_{4}}$ splits into three
dimensional spatial part which evolves upon a one dimensional time-like M$%
_{_{1}}$ part following a trivial fibration

\begin{equation*}
f_{_{1}}:M_{_{4}}\longrightarrow M_{_{1}}
\end{equation*}
with group structure $\frac{G}{SO(3)}$ and M$_{_{1}}$ is the corresponding
invariant part. In this case metrics are of RFW\ type

\begin{equation}
ds^{_{^{^{2}}}}=dt^{^{2}}+a^{^{2}}(t)dl^{^{2}}
\end{equation}
a(t) is the scale of the universe and $dl$ is the spatial linear element.

It is well known that Einstein equations corresponding to the
Hilbert-Einstein action coupled to a scalar field

\begin{equation}
S=\frac{1}{2k}\int\limits_{M_{_{4}}}d^{^{4}}x\left\{ \sqrt{g}\left( R+k\left[
g^{^{\mu \nu }}\partial _{_{\mu }}\phi \partial _{_{\nu }}\phi -2U(\phi )%
\right] \right) \right\}
\end{equation}
reduce in the RFW\ space-time to a simple dynamical system

\begin{eqnarray}
\frac{dH}{dt} &=&kU(\phi )-3H^{^{_{^{2}}}}=V(\phi ,H)\qquad \\
\frac{d\phi }{dt} &=&\pm \sqrt{-\frac{2}{k}V(\phi ,H)}
\end{eqnarray}
with solutions corresponding to expanding, collapsing or stationary
universe, this depend on the content of matter and the initially conditions.
H(t)=$\frac{da}{adt}$ is the Hubble expansion rate. This system obtains from
the action

\begin{eqnarray}
S &=&S_{_{grav}}+S_{_{mat}}\qquad \\
S_{grav} &=&\int\limits_{M_{_{1}}}dt\left\{ -\frac{3}{k}a\left( \frac{da}{dt}%
\right) ^{^{2}}\right\} \qquad \\
S_{mat} &=&\int\limits_{M_{_{1}}}dta^{^{3}}\left\{ \frac{1}{2}\left( \frac{%
d\phi }{dt}\right) ^{^{2}}-U(\phi )\right\} \qquad
\end{eqnarray}

\qquad

To have the complete action we need to add one which governs evolution of $%
\frac{G}{SO(3)}$ upon M$_{1}$or the geodesic flow given by

\begin{equation}
S_{_{geo}}(x,g)=\int\limits_{M_{_{1}}}dt\sqrt{h}h^{^{00}}\partial
_{_{t}}x_{_{i}}\partial _{_{t}}x_{_{j}}G^{^{ij}}\qquad
\end{equation}

So evolution in RFW space-times reduces to a quantum mechanical system

\begin{equation}
S_{_{RFW}}=S_{_{grav}}+S_{_{mat}}+S_{_{geo}}\qquad
\end{equation}
$\ $where the coordinates x$^{^{i}},$ the M$_{_{1}}$-metric h and the scalar
field $\phi $ are the basic fields, the metric g$_{_{ij}},$ the potential
couplings and the metric of the field space look like section upon $\frac{G}{%
SO(3)}$ or field depending couplings. The quantum fluctuation about the
classical solutions obtains from the partition function

\qquad 
\begin{equation}
Z=\sum\limits_{M_{1}-top}\int\limits_{M_{1}}\left[ dh\right] \left[ dx\right]
\left[ d\phi \right] e^{-\left( \frac{i}{h}S_{_{RFW}}\right) }
\end{equation}

This situation, namely reduction of Einstein equation coupled to matter in
RFW universe to dynamical law governing evolution upon the cosmic time $\tau 
$, extend to the spherical case. Four dimensional spherical space-time M$%
_{_{4}}$ admit SO(3) isometry group with its isotropic subgroup SO(2)$\equiv
U(1).\;This$ implies existence of a fibration

\begin{equation}
f_{_{2}}:M_{_{4}}\longrightarrow M_{_{2}}.\qquad
\end{equation}
with structure group SO(3)/U(1) and the base space M$_{_{2}}$ is the
invariant part of M$_{_{4}}$ with respect to SO(3). This fibration is
trivial; this means in particular there exist coordinates system upon which
the M$_{_{4}}$ metric writes

\begin{equation}
ds^{_{2}}=h_{_{\alpha \beta }}d\sigma ^{^{\alpha }}d\sigma ^{^{\beta
}}+g_{_{ij}}dx^{^{i}}dx^{^{j}}
\end{equation}
$\left( h_{_{\alpha \beta }}\right) $ is the metric with the time-like
coordinate $\sigma ^{^{0}}=\tau $ and the spatial one $\sigma ^{^{1}}=\sigma
.$ $\left( g_{_{ij}}\right) $ is the SO(3)/U(1) metric with coordinates (x$%
^{^{2}},$x$^{^{3}}).$ In the commoving frame and spherical coordinates (r,$%
\theta ,\varphi )$ the metric writes

\begin{equation}
ds^{2}=d\tau ^{2}+-e^{\lambda }dr^{2}-e^{2\mu }(d\theta ^{2}+\sin ^{2}\theta
d\varphi ^{2})\qquad
\end{equation}
$\lambda $ and $\mu $ generally depend on (r,$\tau ):$ $\lambda $(r,$\tau )$
et $\mu $(r,$\tau ).$The corresponding Einstein equations with hydrodynamic
energy-momentum tensor reduces to a simple dynamical system on the phase
space with the action

\begin{equation}
S_{gr}(\mu ,\lambda )=\int\limits_{M_{2}}d^{2}\sigma L(\mu ,\dot{\mu}%
,\lambda ,\dot{\lambda})\qquad
\end{equation}

In the following paragraph we derive the explicit form of this action and
point out its symmetries.

The evolution of SO(3)/U(1) orbits upon the base space M$_{2}$ is two
dimensional extension of the above geodesic flow. The natural corresponding
action is SO(3)/U(1) gauge invariant and M$_{2}$ generally covariant. Since
the fibration is trivial the Yang-Mills term vanishes and we get SO(3)/U(1)
WZW model

\begin{equation}
S_{WZW}=\int\limits_{M_{2}}d^{2}\sigma \sqrt{h}h^{\alpha \beta }\partial
_{\alpha }x^{i}\partial _{\beta }x^{j}g_{ij}\qquad
\end{equation}
x$^{i},$ g$_{ij}$ are respectively the SO(3)/U(1) coordinates and metric. $%
\sigma ^{\alpha }$ and $\left( h_{\alpha \beta }\right) $ are their M$_{2}$
analogue. Contribution of fields $\phi _{i}$ gets from the action

\begin{equation}
S_{mat}=\sum\limits_{i}\int\limits_{M_{2}}d^{2}\sigma A_{i}(x)\phi
_{i}(x)\qquad \qquad
\end{equation}
A$_{i}$ are composites of \ SO(3)/U(1) coordinates and their derivatives
with respect to M$_{2}$ ones. Then the evolution of Einstein-matter system
in four dimensional spherical space-time obtains from the action principle

\begin{equation}
S_{2}=S_{gr}+S_{WZW}+S_{mat}
\end{equation}

The quantum fluctuation about its classical solutions get from the partition
function

\begin{equation}
Z_{2}=\sum\limits_{M_{2}-top}\int\limits_{M_{2}}\left[ dh\right] \left[ dx%
\right] \left[ d\phi \right] e^{-\left( \frac{i}{h}S_{2}\right) }\qquad
\end{equation}
which reflects that there is uncertainties only in geometry and topology of M%
$_{2}$ and position in the orbit coordinates.

From the above analysis emerges the following scheme: at fundamental scales
the universe is four dimensional and spherical. Law governing evolution of
matter and geometry upon M$_{2}$ is a formal string theory with the
following correspondences:

$
\begin{array}{cc}
\text{{\large String Framework}} & \text{{\large The present Framework \ }}
\\ 
\text{World-sheet}\sum_{g} & 
\begin{array}{c}
\text{The SO(3) invariantpart of M}_{2}\text{with{\small \ }} \\ 
\text{corrdinates inluding the time variable}
\end{array}
\\ 
\text{The petubative Polyakov Serie} & 
\begin{array}{c}
\text{The partition function of 2D} \\ 
\text{ quantum gravity}
\end{array}
\\ 
\begin{array}{c}
The\text{ action represent a static} \\ 
\text{ two }\dim \text{ensional space-time} \\ 
\text{ sigma model}
\end{array}
& 
\begin{array}{c}
The\text{ action represent fundamental } \\ 
\text{evolution}of\text{ matter in} \\ 
\text{ four dimensional sapce-time.}
\end{array}
\end{array}
$

\bigskip at relatively large scales spherical space-time loses its center,
in particular we reach homogeneity and so fall on RFW universe. This schemes
gives an outcome to the above inadequacies, concerning time dependant
processus in string theory and the interpretation of its building blocks. It
also constitutes a cosmological scenario where we can evaluate quantum
effects in particular when M$_{2}$ =S$^{2},the$ tree level i the string
sense. To carry out this scenario we need first to determine the action S$%
_{grav}.$

\section{Action principle of four dimensional spherical space-time.}

In this paragraph we construct the two dimensional action S$_{grav}(\lambda
,\mu )$ equivalent to Einstein equations coupled to matter

\begin{equation}
R_{\mu \nu }-\frac{1}{2}R=\frac{8\pi G}{C^{4}}T_{\mu \nu }
\end{equation}

in four dimensional spherical space-times in the hydrodynamic case

\begin{eqnarray}
T_{\mu \nu } &=&\left( p+\varepsilon \right) v_{\mu }v_{\nu }+pg_{\mu \nu } 
\notag \\
with\text{ }p &=&k\varepsilon
\end{eqnarray}

and we study its symmetries.

In four dimensional spherical space-times the non vanishing components of
the energy-momentum tensor T$_{\mu \nu }$ are T$_{00},$T$_{11},$T$_{22},$T$%
_{33}$ and T$_{10}$ and the space-time metric writes

\begin{equation}
ds^{2}=d\tau ^{2}+e^{-\lambda }dr^{2}-e^{2\mu }(d\theta ^{2}+\sin ^{2}\theta
d\varphi ^{2})
\end{equation}
T$_{\mu \nu },\nu ,\lambda $ and $\mu $ are (r,$\tau )$ functions.

The Einstein equations are given by

\begin{equation}
\frac{1}{4}e^{-\lambda }(\frac{\mu ^{\prime }}{2}+\mu ^{\prime }\nu ^{\prime
})-e^{-\nu }(\ddot{\mu}-\frac{1}{2}\dot{\mu}\dot{\nu}+\frac{3}{4}\dot{\mu}%
^{2})-e^{-\mu }=\frac{8\pi G}{C^{4}}T_{1}^{1}
\end{equation}

\begin{equation*}
\frac{1}{4}e^{-\nu }(2\nu ^{\prime \prime }+\nu ^{\prime 2}+2u^{\prime
\prime }+u^{\prime 2}\text{ }-u^{\prime }\lambda ^{\prime }-\nu ^{\prime
}\lambda ^{\prime }+u^{\prime }\nu ^{\prime })
\end{equation*}

\begin{equation}
+\frac{1}{4}e^{-\nu }(\dot{\nu}\dot{\lambda}+\dot{\nu}\dot{\mu}-\dot{\mu}%
\dot{\lambda}-2\ddot{\lambda}-\dot{\lambda}^{2}-2\ddot{u}-\dot{u}^{2})=\frac{%
8\pi G}{C^{4}}T_{2}^{2}
\end{equation}

\begin{equation}
-e^{-\lambda }(u^{\prime \prime }+\frac{3}{4}u^{\prime 2}-\frac{u^{\prime
}\lambda ^{\prime }}{4})+\frac{1}{2}e^{-v}+\frac{1}{2}\dot{u}^{2}+e^{-\mu }=-%
\frac{8\pi G}{C^{4}}T_{0}^{0}
\end{equation}

Moreover there is covariant conservation equations

\begin{equation}
\nabla ^{\mu }T_{\mu \nu }=0\qquad
\end{equation}
which implies additional dependance of \ $\mu ,\nu ,\lambda $ and the fields
of which depend T$_{\mu \nu }.$

In the hydrodynamic case conservation equations writes

\begin{eqnarray}
\nu ^{\prime } &=&\frac{2p^{\prime }}{p+\epsilon }\qquad \\
(2\dot{\mu}+\dot{\lambda}) &=&-\frac{2\dot{\epsilon}}{p+\epsilon }\qquad
\end{eqnarray}
these equations and the constraint p=k$\varepsilon $ implies additional
dependences of variables (($\mu ,\nu ,\lambda ,\epsilon ,p).$ We show that
equations($32-34$) are equivalent to equations of motion corresponding to
the action

\begin{equation}
S_{grav}=\int\limits_{M_{2}}L(\lambda ,\mu ,\dot{\mu},\dot{\lambda},\mu
^{\prime },\nu ^{\prime })d\tau dr\qquad
\end{equation}

\begin{equation}
L=L_{1}+L_{2}+L_{3}
\end{equation}

\begin{eqnarray}
L_{1} &=&\left\{ \frac{u^{\prime 2}}{2}+ku^{\prime }(\lambda ^{\prime
}+2u^{\prime }\right\} \exp \left[ -\frac{k+1}{2}\lambda +(k+1)u\right] \\
L_{2} &=&\left\{ \dot{u}^{2}+\dot{\mu}\dot{\lambda}\right\} \exp \left[ -%
\frac{k-1}{2}\lambda +(1-k)u\right] \\
L_{3} &=&2\exp \left[ \frac{k+1}{2}\lambda +ku\right] \qquad
\end{eqnarray}

with the constraints 
\begin{eqnarray}
\tilde{T}_{1}^{0} &=&\dot{\lambda}\frac{\partial L}{\partial \lambda
^{\prime }}+\dot{\mu}\frac{\partial L}{\partial \mu ^{\prime }}=\frac{1}{2}%
e^{-\lambda }\left\{ 2\dot{\mu}^{\prime }+\dot{\mu}\mu ^{\prime }-\dot{%
\lambda}\mu ^{\prime }-\left( k\lambda ^{\prime }+2k\mu ^{\prime }\right) 
\dot{\mu}\right\} =0 \\
\tilde{T}_{0}^{1} &=&\lambda ^{\prime }\frac{\partial L}{\partial \dot{%
\lambda}}+\mu ^{\prime }\frac{\partial L}{\partial \dot{\mu}}=0
\end{eqnarray}

$T_{\alpha \beta }$ is the energy-momentum tensor of $S_{grav}$ (38).

In term of the horizon ''a'', equation of motion take a simple form

\begin{eqnarray}
\dot{a} &=&-\frac{8\pi G}{c^{4}}pr^{2}\dot{r}\qquad \\
a\prime &=&-\frac{8\pi G}{c^{4}}pr^{2}r^{\prime } \\
1-\frac{a}{r} &=&\dot{r}p^{\frac{2k}{k+1}}-r^{\prime }r^{4}\epsilon ^{\frac{2%
}{k+1}}\qquad
\end{eqnarray}

$(47)$ extend the expression of the horizon to non-static case, r=e$^{2\mu }$
is the radius of the spherical system.

In this form, it easy to verify that equations $(45-47)$ are invariant with
respect to the transformations

\begin{eqnarray}
\tau &\longrightarrow &\alpha \tau \\
r &\longrightarrow &\beta r\qquad
\end{eqnarray}
under which the energy density $\epsilon $, the pressure p, the horizon
''a'' and the radius r transform like

\begin{eqnarray}
p &\longrightarrow &\gamma p\qquad \qquad \qquad \epsilon \longrightarrow
\gamma \epsilon \\
r &\longrightarrow &\lambda r\qquad \qquad \qquad a\longrightarrow \lambda
a\qquad \qquad
\end{eqnarray}

Parameters ($\alpha ,\beta ,\gamma ,\lambda )$ verify

\begin{eqnarray}
\alpha ^{2} &=&\lambda ^{-\frac{8k}{k+1}}\beta ^{2} \\
\gamma &=&\lambda ^{-2}\qquad
\end{eqnarray}

This implies in the particular case when the pressure vanishes, K=0,
Einstein equations in four dimensional spherical space-times admits a 2d
conformal symmetry.

\pagebreak

\section{Discussion and conclusion.}

An alternative description of string theory which deals with insufficiency
in our understanding of the string theory is proposed. \emph{\ String theory
with seemingly two dimensional static target space describes in fact
dynamic, including quantum effects, in four dimensional symmetric and
time-dependant geometry. }This permits to deal with the problem of time
dependant solutions of string theory and the inadequate actual
interpretation of its building blocks. The corresponding principle action S$%
_{grav}+S_{WZW}$ + S$_{grav}$ are studied, \emph{the action is conformal
invariant in the pressureless regime.}

The present work should be confronted to works relatives to 't Hooft
proposal on quantization of Back Holes, which 'reconstruct' string paradigm
from quantization of black holes in symmetrical four dimensional space-times
and to the Witten construction of a two dimensional black hole solution $^{%
\left[ 14\right] },$ the corresponding action consist of the $\frac{SL(2R)}{%
U(1)}\equiv SU(1,1)$ WZW model and additional dilaton coupling \emph{which
we get from the perturbative quantum correction.} In the first order this
coupling writes 
\begin{equation*}
\int\limits_{M_{2}}\phi (r)R^{(2)}\sqrt{h}
\end{equation*}
$\phi (r)$ is the dilaton field, some function of $SU(1,1)$ coordinates, and 
$R^{(2)}$ is the curvature of the world-sheet metric of M$_{2}.$ Moreover
this model admits W$_{\infty }$ symmetries$^{\left[ 15\right] }.$ there are
similarities that shares this model with S$_{grav}+S_{WZW}.$ The $\frac{SO(3)%
}{U(1)}$ WZW model up a Z$_{2}$ factor, looks like an Euclidean continuation
of $\frac{SL(2R)}{U(1)}$ WZW model. S$_{grav}$ is comparable to the dilaton
coupling with r = e$^{2\mu },$ the world-sheet M$_{2}$ coincides with the
invariant part of the four dimensional spherical space-time with respect to
SO(3). Moreover S$_{grav}+S_{WZW}$ should admit the above W$_{\infty }$
(this is the case for the $\frac{SO(3)}{U(1)}$ WZW model). Equations of
motion corresponding to S$_{grav}$ constitute a quasi-homogeneous system
(similitude) with Weight depending on the ratio k of pressure to the energy
density. It is tempting to show that similitude admits W$_{\infty }$
symmetries.

Following the above work of Witten, identification of S$_{grav}$ to dilaton
coupling is likely to carry out the induced gravity idea in four dimensional
space-times$^{\left[ 16\right] }.$ It is known that this coupling is induced
by perturbative quantum corrections. This implies that fundamental evolution
in four dimensional spherical space-times is given \emph{at the classical
level} by the $\frac{SO(3)}{U(1)}$ WZW model. The Hilbert-Einstein action
coupled to matter, which is equivalent to S$_{grav}$, get in fact from
quantum corrections.\pagebreak

\bigskip

\bigskip {\LARGE \vspace*{0.23cm}\bigskip REFE\smallskip RENCES.\linebreak }

$[1]$P.Langacker Phys.Rep {\small 72(1981)56.}and references therein.

$[2]$A.Lindei ``Inflation and Quantum Cosmoogy'' CERN-TH {\small 5561/89}%
.and references therein.

$[3]$M.Green, J.Schwarz and E.Witten ``Superstrig Theory''Volumes I and II
Cambridge University Press{\small \ 1987.}

$[4]$A.sen hep-th:9802051 and references therein.

$[5]$G 't Hooft ''Dimensional Reduction in Quantum Gravity'' gr-qc/{\small %
9310026.}

L.Susskind ''The World as a Hologram'' hep-th/{\small 9409089}. J.Math.Phys 
{\small 36(1995)6377}.

$[6]$J.Maldacena The large N limit of Superconformal field Theories and
Supergravity'' hep-th/{\small 9711200.}Adv.theor.Math.Phys {\small %
2(1998)231.}

$[7]$G. 't Hooft ``Constructive quantization of Black Holes'' Lectures at
Spring School of string theory, gauge theory and quantum gravity{\small \
19-27} April {\small 1993 }ICTP Trieste Italy.

\ \ \ G. 't Hooft `` Black Holes and quantum mechanics'' Acta Physica
Polinica Vol B{\small 19 (1988)187.}

$[8]$C.Thorn phys. report {\small (1989).}

$[9]$E.Witten ''the search for higher symetry in string theory'' lectures at
th discussion meeting on string theory of the royal society, London,
December {\small 1998.}

$[10]$G.T. Horowitz and R.Steif Phys. Rev. lett {\small 64 (1990)260D42
(1990) 1950.}

$[11]$Justin Khoury and al ``From Big Crunch to Big Bang'' hep-th/{\small %
0108187 v4 \ 2 Mar 2002.}

\ \ \ Hong Liu et al `` St rings in a time-dependant orbifold'' hep-th/%
{\small 0204168 v3 7June 2002.}

$[12]$C.Nappi ICTP Preprints{\small \ 1989.}

$[13]$D.Gepner and E.Witten Nucl.Phys B {\small (1986).}

$[14]$E.Witten Phys.Rev D {\small 44 (1991)314.}

$[15]$S.Chandhuri et al ''Steing theory, black holes and SL(2R) Current
Algabra'' Fermi Pup{\small -92/69-T.}

$[16]$A.D Sakharov Sov.phys.Dok{\small . 12(1968)1040.}

\bigskip

\bigskip

\bigskip

$\mathstrut \bigskip $

\end{document}